\documentclass{article}

\usepackage{iclr2027_conference,times}

\usepackage{amsmath,amsfonts,bm}

\def\eqref#1{equation~\ref{#1}}

\def\1{\bm{1}}

\DeclareMathAlphabet{\mathsfit}{\encodingdefault}{\sfdefault}{m}{sl}
\SetMathAlphabet{\mathsfit}{bold}{\encodingdefault}{\sfdefault}{bx}{n}

\usepackage{amsmath,amssymb,mathtools}
\usepackage{booktabs}
\usepackage{graphicx}
\usepackage{microtype}
\usepackage{hyperref}
\usepackage{xurl}
\hypersetup{
    pdftitle={CONTRA: Discovering and Qualifying Behavior-Changing Questions for Selective Clarification in LLM Code Generation},
    pdfauthor={Zheng Fang, Yongmin Li, Yichang Zhang, Dongming Jin, Haoyu Wang, Shuai Wang, Zhi Jin, Ge Li}
}

\usepackage{multirow}
\usepackage{capt-of}
\usepackage{wrapfig}

\newcommand{\model}{\textsc{Contra}}

\title{\raggedright CONTRA: Discovering and Qualifying Behavior-Changing Questions for Selective Clarification in LLM Code Generation}

\author{
\textbf{Zheng Fang$^{1}$, Yongmin Li$^{1}$, Yichang Zhang$^{1}$, Dongming Jin$^{1}$} \\
\textbf{Haoyu Wang$^{1}$, Shuai Wang$^{1}$, Zhi Jin$^{2}$, Ge Li$^{1}$}\thanks{Corresponding author: Ge Li (\texttt{lige@pku.edu.cn}).} \\
$^{1}$Peking University \qquad $^{2}$Wuhan University \\
\texttt{fangz@pku.edu.cn} \qquad \texttt{lige@pku.edu.cn}
}

\iclrfinalcopy

\begin{document}
\raggedbottom

\maketitle
\fancyhead{}

\begin{abstract}

Coding agents can generate code that appears correct but implements behavior the user never intended. This mismatch can arise when an agent silently resolves underspecified requirements through its own assumptions. As subsequent development builds on these assumptions, correcting the resulting behavior can become increasingly costly. Early clarification can help prevent such mismatches, but unnecessary questions can interrupt developers and slow down development. Existing methods struggle to identify key clarification questions while avoiding unnecessary ones. Therefore, we propose \model{}, a training-free method that combines broad question discovery with semantic and execution-based question qualification. \model{} first generates candidate questions and filters out those unrelated to required behavior or already resolved by the requirement. For each remaining question, it generates programs conditioned on two plausible answers and checks for stable behavioral differences on shared inputs. It then uses the interaction history to select among qualified questions or stop asking. Experiments on ClarifyCodeBench show that \model{} achieves the highest F1 with all four coding agents, exceeding the best baseline macro-average F1 by 13.88 percentage points. With the same LLM and evaluation protocol, \model{} also achieves higher clarification recall and F1 than the coding harnesses Claude Code and OpenHands.
 To support practical use, we also implement \model{} as a Claude Code plugin that integrates selective clarification into everyday development.\footnote{Released method components and plugin: \url{https://github.com/fangz-cs/Contra}.}

\end{abstract}

\section{Introduction}
\label{sec:introduction}

Coding agents are increasingly used to generate and modify code from natural-language requirements~\citep{metagpt, openhands, swe-agent}. However, these requirements often leave important aspects of program behavior underspecified, such as how to handle edge cases or format outputs~\citep{fakhoury2024llm, clarifygpt, clarifycodebench}. Faced with such requirements, an agent may silently adopt one plausible interpretation and proceed with implementation. The resulting code may be correct under that interpretation, yet fail to match the user's intended behavior~\citep{fakhoury2024llm, ambig-swe, ask-or-assume}. If this mismatch goes unnoticed, subsequent development may build on the same assumptions, making correction increasingly costly. Therefore, clarification helps make these assumptions explicit and resolve the underlying choices with the user before the code becomes the basis for further development.

However, excessive questioning can impose unnecessary interaction costs on developers~\citep{zhang2025clarify,vijayvargiya2026asking}.
For example, an agent may ask questions already answered by the requirements or earlier interactions, or raise questions unrelated to the task.
This motivates selective clarification, where an agent seeks additional user input only when needed~\citep{clam}.
For coding agents, this requires identifying necessary questions within a task while avoiding redundant or irrelevant ones.

Our results in Table~\ref{tab:main-results} illustrate this trade-off. Direct Prompting~\citep{ambig-swe, clarifycodebench} and Ask-or-Assume~\citep{ask-or-assume} rely on the model to decide whether to ask, but often miss necessary questions. By comparison, ClarifyGPT~\citep{clarifygpt} generates questions from behavioral disagreement among sampled programs, but asks many unnecessary questions. Such disagreement can reflect implementation errors even when the requirements are clear. Therefore, selective clarification requires both broad question discovery and careful qualification.

To address this challenge, we propose \model{}, which combines broad question discovery with careful qualification. It first discovers candidate questions and checks whether each is relevant to the requirement and remains unresolved. For each remaining question, \model{} generates programs conditioned on two plausible answers and checks for stable behavioral differences on shared inputs. During interaction, it uses the user's answers to select the next qualified question or stop asking. This design identifies key clarification questions while filtering out unnecessary ones.

We evaluate \model{} on ClarifyCodeBench~\citep{clarifycodebench} with coding agents built on GPT-5-mini~\citep{gpt5-system-card}, GPT-5.5~\citep{gpt55-system-card}, Grok~4.5~\citep{grok45}, and Gemini~3.1 Pro~\citep{gemini31pro}.
Under the same five-question budget, we compare \model{} with Direct Prompting, Ask-or-Assume~\citep{ask-or-assume}, and ClarifyGPT~\citep{clarifygpt}.
\model{} achieves the highest F1 and turn-discounted key question rate (TKQR) for all four agents.
Its macro-average F1 and TKQR reach 41.20\% and 38.32\%, exceeding the best baseline for each metric by 13.88 and 12.33 percentage points, respectively.
Furthermore, using Qwen3.8-27B~\citep{qwen38-27b} and the same evaluation protocol, \model{} achieves higher recall and F1 than Claude Code~\citep{claude-code} and OpenHands~\citep{openhands}.
Its F1 reaches 19.86\%, compared with 9.63\% for Claude Code and 10.61\% for OpenHands.
Finally, we integrate \model{} into Claude Code as a plugin that identifies and reports implementation decisions requiring user clarification.

Our main contributions are as follows.
\begin{itemize}
    \item We propose \model{}, a selective clarification method that combines broad question discovery, semantic and execution-based question qualification, and adaptive question selection. Its execution-based qualification links candidate questions to behavioral differences between programs generated under alternative answers.

    \item We demonstrate consistent improvements on ClarifyCodeBench across four coding agents, achieving the highest F1 and TKQR among the compared methods. Under the same LLM and evaluation protocol, \model{} also achieves higher clarification recall and F1 than the coding harnesses Claude Code and OpenHands.

    \item We implement \model{} as a Claude Code plugin that identifies and reports implementation decisions requiring clarification, helping developers confirm intended program behavior during everyday development.
\end{itemize}

\section{Related Work}
\label{sec:related_work}

\paragraph{Requirement clarification and specification refinement.}
Prior work improves LLM code generation by making user intent more explicit. CodeClarQA~\citep{codeclarqa} provides questions and answers that enrich underspecified requirements for code generation. TiCoder~\citep{fakhoury2024llm} elicits user feedback on generated tests and uses that feedback to formalize intent and rank candidate programs. ClarifyCoder~\citep{clarifycoder} uses synthetic data and instruction tuning to teach code LLMs to recognize underspecification and request clarification. Other methods refine specifications without user interaction. SpecFix~\citep{specfix} analyzes the behavioral distribution of sampled programs and maps changes in that distribution back to a repaired problem description. Specine~\citep{specine} identifies mismatches between an input specification and the specification inferred from generated code, then constructs an aligned specification for subsequent generation. While these methods refine requirements, \model{} qualifies individual questions by checking relevance, whether they remain unresolved, and whether plausible answers change program behavior.

\paragraph{Selective clarification.}
Selective clarification concerns deciding when additional user input is necessary. CLAM~\citep{clam} studies whether an LLM should answer directly or request clarification, while IntentSim~\citep{zhang2025clarify} estimates clarification utility from uncertainty over simulated user intents. Learning-based approaches use simulated future turns to label preferences~\citep{zhang2025modeling}, or train clarification policies using task relevance and user answerability~\citep{vijayvargiya2026asking}. In code generation, ClarifyGPT~\citep{clarifygpt} uses behavioral disagreement among programs sampled from the same requirement to decide whether to clarify. Ask-or-Assume~\citep{ask-or-assume} uses the coding agent or a separate intent agent to decide when user input is needed. \model{} evaluates each candidate question through semantic checks and behavioral verification, then uses the interaction history to select qualified questions that still require user input.

\section{Methodology}
\label{sec:method}

\subsection{Problem Formulation}
\label{sec:problem-formulation}

Let \(S\) denote a natural-language programming requirement. If \(S\) is underspecified, a clarification question \(q\) requests the additional user input needed to implement the intended behavior.

Given \(S\) and a question budget \(B\), the agent may ask at most \(B\) questions. At step \(t\), the interaction history contains all previous questions and user answers,
\[
H_t = \bigl[(q_1,a_1),\ldots,(q_{t-1},a_{t-1})\bigr],
\]
where \(a_i\) is the user's answer to question \(q_i\). Based on \(S\) and \(H_t\), the agent either asks another question or stops. The objective is to identify and ask key clarification questions while avoiding unnecessary ones, including questions unrelated to the task or already answered by \(S\) or \(H_t\).

\subsection{Overview}
\label{sec:method-overview}

\begin{figure}[!htbp]
    \centering
    \includegraphics[width=\textwidth]{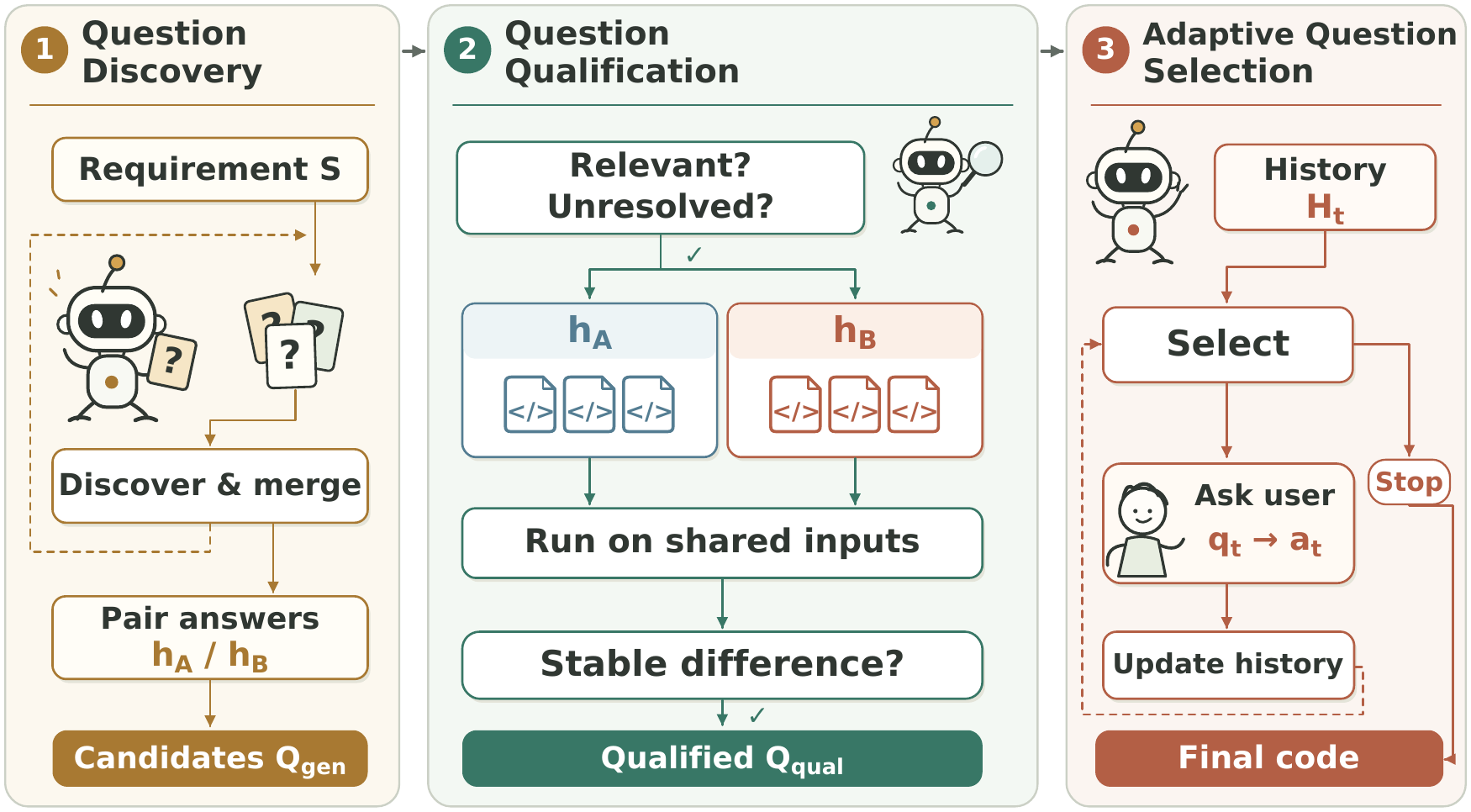}
    \caption{Overview of \model{} for selective clarification.}
    \label{fig:method-overview}
\end{figure}

Figure~\ref{fig:method-overview} shows the three stages of \model{}. Question Discovery constructs a broad candidate set \(Q_{\mathrm{gen}}\). Question Qualification applies semantic checks and behavioral verification to obtain \(Q_{\mathrm{qual}}\). Finally, Adaptive Question Selection uses the history to select a qualified question or stop.

\subsection{Question Discovery}
\label{sec:question-discovery}

Question Discovery generates a broad set of candidate clarification questions from \(S\). Each candidate includes a question and two plausible answers, which support candidate review and merging.

To expand the candidate set, each discovery round combines context-aware and independent generation. Context-aware generation uses \(S\) and the current candidate set to search for additional questions, while independent generation uses only \(S\). The new and existing candidates are jointly reviewed to remove duplicates and merge overlapping questions. Discovery stops after two consecutive rounds add no distinct questions or a predefined round limit is reached. After discovery, \model{} regenerates two plausible answers for each retained question \(q\) as requirement completions \(h_A\) and \(h_B\). The completions should be compatible with \(S\) and differ only in their answer to \(q\). This constraint links subsequent behavioral differences to the question being evaluated and avoids introducing unrelated assumptions. Questions with completion pairs form \(Q_{\mathrm{gen}}\) for Question Qualification.
 
\subsection{Question Qualification}
\label{sec:question-qualification}

Question Qualification evaluates each candidate question together with its completion pair. It retains questions that are relevant, remain unresolved by \(S\), and pass behavioral verification. The qualified set is
\[
Q_{\mathrm{qual}}
=
\left\{
q \in Q_{\mathrm{gen}} \mid
\mathrm{Rel}(q)
\land \mathrm{Unres}(q)
\land \mathrm{Diff}(q)
\right\}.
\]

\paragraph{Relevance and Resolution.}
Relevance checks whether \(q\) concerns required program behavior and whether its completions specify different alternatives for that behavior. Resolution then checks whether \(S\) already determines the answer. A question is marked as resolved only when text in \(S\) supports one completion and rules out the other. The judgment must cite this supporting text. Only relevant and unresolved questions proceed to behavioral verification.

\paragraph{Behavioral Verification.}
For each completion \(h_z\), where \(z \in \{A,B\}\), \model{} independently generates three programs \(p_z^1,p_z^2,p_z^3\) conditioned on \(S\), \(q\), and \(h_z\). It also generates a shared set \(\mathcal{X}\) of at least seven inputs valid under both completions using \(S\), \(q\), \(h_A\), and \(h_B\).

All six programs are executed on the shared inputs. For each completion, the output on input \(x\) is considered stable when at least two programs agree,
\[
o_z(x)=y
\quad\Longleftrightarrow\quad
\left|
\left\{j \in \{1,2,3\} \mid p_z^j(x)=y\right\}
\right| \geq 2.
\]
If no output receives at least two votes, \(o_z(x)\) is undefined. The question passes behavioral verification when the two groups produce different stable outputs on at least one shared input,
\[
\begin{aligned}
\mathrm{Diff}(q)=1
\quad\Longleftrightarrow\quad
&\exists x \in \mathcal{X} \text{ such that}\\
&o_A(x),o_B(x) \text{ are defined and } o_A(x)\neq o_B(x).
\end{aligned}
\]
Otherwise, \(\mathrm{Diff}(q)=0\), and the question is discarded.

This comparison tests the consequences of alternative answers to a specific question. Shared inputs make the two groups directly comparable, while agreement within each group reduces the influence of an isolated implementation error. A question is retained only when execution provides evidence of a behavioral difference. A failed check does not establish behavioral equivalence.

\subsection{Adaptive Question Selection}
\label{sec:adaptive-question-selection}

A user response may resolve another qualified question or make it unnecessary. Therefore, Adaptive Question Selection uses the interaction history to reassess which questions still require user input.

Let \(Q_t \subseteq Q_{\mathrm{qual}}\) denote the remaining questions at step \(t\), with \(Q_1=Q_{\mathrm{qual}}\). Given \(S\), \(H_t\), and the text of the questions in \(Q_t\), the selector either chooses a question that still requires user input or returns \textsc{Stop} when further clarification is unnecessary,
\[
\mathrm{Select}(S,H_t,Q_t)
\in Q_t \cup \{\text{\textsc{Stop}}\}.
\]

After receiving \(a_t\), \model{} appends \((q_t,a_t)\) to the history and removes \(q_t\) from the candidate set. The updated history helps avoid questions that have become resolved or redundant. Clarification ends when the selector returns \textsc{Stop}, no questions remain, or \(B\) questions have been asked.

\section{Experiments}
\label{sec:experiments}

\begin{table}[!t]
    \centering
    \caption{Clarification results (\%) on ClarifyCodeBench. P and R denote precision and recall. Bold values indicate the best result for each metric and coding agent.}
    \label{tab:main-results}
    \small
    \setlength{\tabcolsep}{2pt}
    \renewcommand{\arraystretch}{1.05}
    \medskip

    \begin{minipage}[t]{0.485\textwidth}
        \centering
        \begin{tabular*}{\linewidth}{@{\extracolsep{\fill}}lrrrr@{}}
            \toprule
            \multicolumn{5}{c}{\textbf{GPT-5-mini}} \\
            \midrule
            Method & F1 $\uparrow$ & TKQR $\uparrow$
            & P $\uparrow$ & R $\uparrow$ \\
            \midrule
            Direct Prompting
            & 16.86 & 15.09 & 20.62 & 14.25 \\
            Ask-or-Assume
            & 17.06 & 17.00 & 17.06 & 17.06 \\
            ClarifyGPT
            & 19.92 & 30.94 & 13.58 & \textbf{37.37} \\
            \textbf{\model{}}
            & \textbf{35.36} & \textbf{34.32}
            & \textbf{36.20} & 34.56 \\
            \bottomrule
        \end{tabular*}
    \end{minipage}
    \hfill
    \begin{minipage}[t]{0.485\textwidth}
        \centering
        \begin{tabular*}{\linewidth}{@{\extracolsep{\fill}}lrrrr@{}}
            \toprule
            \multicolumn{5}{c}{\textbf{Grok 4.5}} \\
            \midrule
            Method & F1 $\uparrow$ & TKQR $\uparrow$
            & P $\uparrow$ & R $\uparrow$ \\
            \midrule
            Direct Prompting
            & 23.12 & 17.24 & \textbf{41.81} & 15.98 \\
            Ask-or-Assume
            & 29.17 & 25.54 & 37.25 & 23.97 \\
            ClarifyGPT
            & 20.90 & 27.83 & 16.16 & 29.59 \\
            \textbf{\model{}}
            & \textbf{36.54} & \textbf{35.95}
            & 36.58 & \textbf{36.50} \\
            \bottomrule
        \end{tabular*}
    \end{minipage}

    \par\medskip

    \begin{minipage}[t]{0.485\textwidth}
        \centering
        \begin{tabular*}{\linewidth}{@{\extracolsep{\fill}}lrrrr@{}}
            \multicolumn{5}{c}{\textbf{Gemini 3.1 Pro}} \\
            \midrule
            Method & F1 $\uparrow$ & TKQR $\uparrow$
            & P $\uparrow$ & R $\uparrow$ \\
            \midrule
            Direct Prompting
            & 37.12 & 33.93 & 44.68 & 31.75 \\
            Ask-or-Assume
            & 37.55 & 35.54 & 43.47 & 33.05 \\
            ClarifyGPT
            & 17.14 & 18.64 & 16.60 & 17.71 \\
            \textbf{\model{}}
            & \textbf{50.49} & \textbf{45.37}
            & \textbf{58.74} & \textbf{44.28} \\
            \bottomrule
        \end{tabular*}
    \end{minipage}
    \hfill
    \begin{minipage}[t]{0.485\textwidth}
        \centering
        \begin{tabular*}{\linewidth}{@{\extracolsep{\fill}}lrrrr@{}}
            \multicolumn{5}{c}{\textbf{GPT-5.5}} \\
            \midrule
            Method & F1 $\uparrow$ & TKQR $\uparrow$
            & P $\uparrow$ & R $\uparrow$ \\
            \midrule
            Direct Prompting
            & 24.10 & 16.29 & \textbf{59.32} & 15.12 \\
            Ask-or-Assume
            & 25.51 & 21.64 & 34.31 & 20.30 \\
            ClarifyGPT
            & 23.53 & 26.53 & 20.03 & 28.51 \\
            \textbf{\model{}}
            & \textbf{42.39} & \textbf{37.63}
            & 50.15 & \textbf{36.72} \\
            \bottomrule
        \end{tabular*}
    \end{minipage}
\end{table}

\subsection{Experimental Setup}
\label{sec:experimental-setup}

\paragraph{Benchmark and Evaluation.}
We evaluate on ClarifyCodeBench~\citep{clarifycodebench}, which provides underspecified programming requirements with gold clarification questions and answers. Our evaluation uses 419 underspecified tasks and the 80 fully specified tasks provided by ClarifyCodeBench. Each method may ask at most five clarification questions before generating a final program. The benchmark's GPT-4o evaluator matches questions against each task's gold questions by majority vote over three judgments.
\paragraph{Baselines.}
We compare \model{} with three training-free baselines. Direct Prompting uses the benchmark prompt to ask a question or generate code. Ask-or-Assume~\citep{ask-or-assume} adds an intent agent that reviews the proposed action and requests clarification when needed. ClarifyGPT~\citep{clarifygpt} generates questions from behavioral disagreement among programs sampled from the current requirement and executed on shared inputs. We also compare with the coding harnesses Claude Code~\citep{claude-code} and OpenHands~\citep{openhands}.
\paragraph{Evaluation Metrics.}
We report question precision, recall, F1, and turn-discounted key question rate (TKQR)~\citep{clarifycodebench}. Precision is the proportion of all questions asked that match a previously uncovered gold question, including questions asked on fully specified tasks in the denominator. Recall is the proportion of gold questions matched by at least one question, and F1 is the harmonic mean of precision and recall. TKQR rewards earlier identification of gold questions using normalized discounted cumulative gain~\citep{jarvelin2002cumulated}, averaged across tasks. For the harness comparison, we also report the false-positive rate (FPR): the proportion of fully specified tasks on which at least one question is asked.
\paragraph{Models and Implementation.}
We evaluate coding agents built on GPT-5-mini, GPT-5.5, Grok~4.5, and Gemini~3.1 Pro. For each comparison, all components of \model{} and the baselines use the same LLM. We use low reasoning effort, with stochastic sampling only for program generation in Behavioral Verification and ClarifyGPT. The harness comparison uses Qwen3.8-27B for \model{}, Claude Code, and OpenHands, with the same question budget and scoring procedure. We execute generated Python programs in isolated environments under fixed time, memory, and output limits.

\subsection{Main Results}
\label{sec:main-results}

\begingroup
\setlength{\intextsep}{0pt}
\begin{wraptable}{R}{0.49\textwidth}
\centering
    \setlength{\abovecaptionskip}{0pt}
    \caption{Macro-average results (\%) across the four agents. Bold values indicate the best result.}
    \label{tab:macro-results}
    \small
    \setlength{\tabcolsep}{2pt}
    \renewcommand{\arraystretch}{1.05}
    \medskip

    \begin{tabular*}{\linewidth}{@{\extracolsep{\fill}}lrrrr@{}}
        \toprule
        Method & F1 $\uparrow$ & TKQR $\uparrow$
        & P $\uparrow$ & R $\uparrow$ \\
        \midrule
        Direct Prompting
        & 25.30 & 20.64 & 41.61 & 19.28 \\
        Ask-or-Assume
        & 27.32 & 24.93 & 33.02 & 23.60 \\
        ClarifyGPT
        & 20.37 & 25.99 & 16.59 & 28.29 \\
        \textbf{\model{}}
        & \textbf{41.20} & \textbf{38.32}
        & \textbf{45.42} & \textbf{38.01} \\
        \bottomrule
    \end{tabular*}
\end{wraptable}
\model{} achieves the highest F1 and TKQR for every coding agent (Table~\ref{tab:main-results}). Its macro-average F1 and TKQR reach 41.20\% and 38.32\%, exceeding the best baselines by 13.88 and 12.33 percentage points, respectively. Table~\ref{tab:macro-results} further shows that \model{} achieves the highest macro-average precision and recall. These gains reflect more key questions identified and a lower proportion of unnecessary questions.

The baseline results illustrate the difficulty of identifying
key questions while avoiding unnecessary ones. With GPT-5.5,
Direct Prompting achieves 59.32\% precision but only 15.12\% recall,
leaving many key questions unasked. \model{} increases recall
to 36.72\% while maintaining 50.15\% precision. Conversely,
with GPT-5-mini, ClarifyGPT achieves 37.37\% recall but only
13.58\% precision, indicating that many questions do not
identify a previously uncovered gold question. Here,
\model{} raises precision to 36.20\% while retaining a
comparable recall of 34.56\%. Despite slightly lower recall, \model{} improves TKQR from 30.94\% to 34.32\%, indicating that it finds key questions earlier and uses clarification turns more efficiently.

The relative performance of baseline methods also varies
across coding agents. For example, ClarifyGPT achieves the
highest baseline recall with GPT-5-mini but the lowest with
Gemini 3.1 Pro. This variation suggests that its effectiveness
depends strongly on the coding agent. By comparison,
\model{} consistently achieves the highest F1 and TKQR
across all four agents, showing that the gains hold across
the evaluated coding agents.

\par
\endgroup

\begin{figure}[!htbp]
    \centering
    \includegraphics[width=\textwidth]{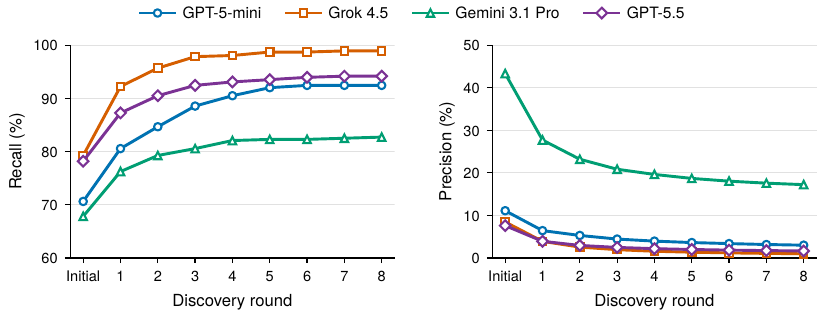}
    \caption{Recall and precision across discovery rounds. Each point evaluates all distinct candidates found up to that round. Initial denotes the first generation step.}
    \label{fig:qg-refinement}
\end{figure}

\subsection{Effect of Discovery Rounds}
\label{sec:qg-analysis}

Question Discovery achieves high recall across all four
coding agents, recovering most gold clarification questions
within approximately four discovery rounds.
Figure~\ref{fig:qg-refinement} shows how recall and precision
change as candidate questions accumulate. Recall rises rapidly in early rounds before saturating, indicating that iterative discovery finds key questions missed initially.

However, precision declines as the candidate set grows.
This indicates that additional rounds also introduce many
questions that do not identify additional gold questions.
Therefore, broad discovery provides a candidate set with high recall,
while Question Qualification filters unnecessary questions from this set.

\begin{figure}[!htbp]
    \centering
    \includegraphics[width=\textwidth]{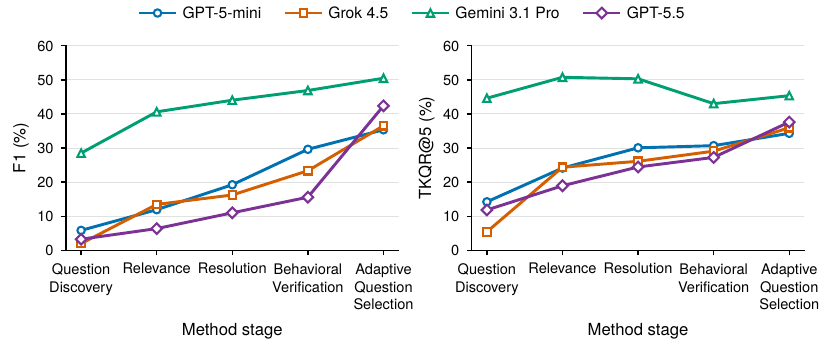}
    \caption{F1 and TKQR@5 across \model{} stages. Before Adaptive Question Selection, F1 evaluates all retained candidates and TKQR@5 evaluates the first five questions in discovery order.}
    \label{fig:qualification-stages}
\end{figure}

\subsection{Effect of Question Qualification}
\label{sec:qualification-analysis}

Figure~\ref{fig:qualification-stages} shows that F1 increases
at every qualification stage across all four coding agents.
Relevance and Resolution improve candidate
quality, and Behavioral Verification provides further gains.
These consistent improvements demonstrate the effectiveness
of successive filtering in balancing the retention of key
questions with the removal of unnecessary ones.

Filtering alone also improves TKQR. Across the four coding agents, macro-average
TKQR increases from 19.03\% after Question Discovery to
32.52\% after Behavioral Verification. The qualification
checks only remove candidates and preserve the relative
order of the remaining questions. Removing
unnecessary questions allows key questions to appear earlier
in the sequence. Beyond improving candidate quality, successive filtering thus brings key questions forward, making more effective use of the question budget.

\subsection{Effect of Adaptive Question Selection}
\label{sec:adaptive-selection-analysis}

Adaptive Question Selection uses the interaction history to avoid asking qualified questions that user answers have already resolved or made unnecessary.

The final transition in Figure~\ref{fig:qualification-stages}
shows that Adaptive Question Selection further improves
both F1 and TKQR across all four coding agents.
Macro-average precision increases from 21.57\% to 45.42\%,
while recall decreases from 50.16\% for the full qualified
candidate set to 38.01\% for the questions actually asked.
This transition compares the full candidate set with a five-question interaction. The resulting macro-average F1 increases from 28.86\% to
41.20\%, and TKQR rises from 32.52\% to 38.32\%.
These gains show more efficient identification of key questions within the five-question budget.

\subsection{Ablation Analysis}
\label{sec:ablation-analysis}

\nointerlineskip
\noindent
\begin{minipage}[t]{0.48\textwidth}
\vspace{0pt}
\setlength{\parskip}{0pt}
Table~\ref{tab:gpt5mini-component-ablation} shows that removing any component reduces F1 for GPT-5-mini. The full method achieves the highest precision and F1.
\par\smallskip
Relevance checks required behavior, while Resolution checks whether the requirement already supplies an answer. Removing Relevance lowers precision from 36.20\% to 31.27\%, while removing Resolution lowers it to 26.46\%. The larger decline highlights the need to avoid asking for information already supplied in the requirement.
\end{minipage}\hfill
\begin{minipage}[t]{0.49\textwidth}
\vspace{0pt}
\centering
    \setlength{\abovecaptionskip}{0pt}
    \captionof{table}{Component ablations (\%) with GPT-5-mini. Bold marks the best result for each metric.}
    \label{tab:gpt5mini-component-ablation}
    \small
    \setlength{\tabcolsep}{2pt}
    \renewcommand{\arraystretch}{1.08}
    \smallskip

    \begin{tabular*}{\linewidth}{@{\extracolsep{\fill}}lrrrr@{}}
        \toprule
        Variant & F1 $\uparrow$ & TKQR $\uparrow$
        & P $\uparrow$ & R $\uparrow$ \\
        \midrule
        \textbf{\model{}}
        & \textbf{35.36} & 34.32 & \textbf{36.20} & 34.56 \\
        w/o REL & 32.54 & 32.91 & 31.27 & 33.91 \\
        w/o RES & 29.13 & 32.98 & 26.46 & 32.40 \\
        w/o BV  & 32.58 & \textbf{37.10}
                & 28.04 & \textbf{38.88} \\
        w/o AQS & 26.69 & 30.70 & 20.96 & 36.72 \\
        \bottomrule
    \end{tabular*}

    \smallskip
    {\footnotesize\raggedright
    REL, Relevance; RES, Resolution; BV, Behavioral Verification;
    AQS, Adaptive Question Selection.\par}
\end{minipage}
\par

Behavioral Verification further improves precision. Removing it increases recall and TKQR, but lowers precision from 36.20\% to 28.04\% and F1 from 35.36\% to 32.58\%. This trade-off supports checking how candidate answers affect program behavior before retaining questions.

Removing Adaptive Question Selection causes the largest drop
in precision and F1. F1 falls by 8.67 percentage
points, while TKQR decreases from 34.32\% to 30.70\%, despite
higher recall. These results highlight the value of interaction history in selecting useful questions within a limited budget.

Removing Behavioral Verification and removing Adaptive Question Selection both increase recall, but only the former improves TKQR. Therefore, finding more key questions does not by itself ensure that they are asked early. This contrast highlights their different effects on question timing.



\section{Practical Evaluation and Deployment}
\subsection{Effect on Downstream Code Generation}
\label{sec:downstream-code-generation}

Each agent generates a final program from the original underspecified requirement and the complete clarification history. We evaluate these programs using the benchmark test cases.

\noindent
\begin{minipage}[t]{0.48\textwidth}
    \vspace{0pt}
    \setlength{\parskip}{0.5\baselineskip}

    Table~\ref{tab:main-pass1} reports the highest pass@1 for \model{} with both agents on underspecified tasks. With Gemini~3.1 Pro, it reaches 57.32\%, exceeding Direct Prompting by 1.22 percentage points. With GPT-5.5, it reaches 64.63\%, exceeding ClarifyGPT by 0.61 percentage points. These modest gains across both agents suggest that effective clarification can improve program correctness.
\end{minipage}\hfill
\begin{minipage}[t]{0.49\textwidth}
    \vspace{0pt}
    \centering
    \setlength{\abovecaptionskip}{0pt}
    \captionof{table}{Downstream pass@1 (\%) on ClarifyCodeBench. Bold marks the best result.}
    \label{tab:main-pass1}
    \small
    \setlength{\tabcolsep}{2pt}
    \renewcommand{\arraystretch}{1.08}
    \medskip

    \begin{tabular*}{\linewidth}{@{\extracolsep{\fill}}lrr@{}}
        \toprule
        Method & Gemini 3.1 Pro & GPT-5.5 \\
        \midrule
        Direct Prompting   & 56.10 & 62.50 \\
        Ask-or-Assume & 55.49 & 63.72 \\
        ClarifyGPT    & 53.35 & 64.02 \\
        \midrule
        \textbf{\model{}}
        & \textbf{57.32} & \textbf{64.63} \\
        \bottomrule
    \end{tabular*}
\end{minipage}
\par

Clarification quality measures whether questions recover the intended requirements; pass@1 additionally measures whether the agent implements those requirements correctly. Reporting both captures these two stages of the interaction.

\subsection{Comparison with Coding Harnesses}
\label{sec:harness}

We compare \model{} with Claude Code and OpenHands on ClarifyCodeBench.
All three use Qwen3.8-27B under the same question budget and evaluation
protocol.

\noindent
\begin{minipage}[t]{0.48\textwidth}
\vspace{0pt}
\setlength{\parskip}{0.5\baselineskip}
Table~\ref{tab:harness} shows that \model{} achieves the highest F1 and
TKQR, reaching 19.86\% and 18.79\%, respectively.
These exceed the best harness results by 9.25 and 9.80 percentage points.
Its recall reaches 20.73\%, compared with 6.26\% for Claude Code and 8.42\% for OpenHands, indicating that more key clarification questions are identified under the same budget.
\end{minipage}\hfill
\begin{minipage}[t]{0.49\textwidth}
\vspace{0pt}
\setlength{\abovecaptionskip}{0pt}
\captionof{table}{Clarification results (\%) with Qwen3.8-27B.
Bold marks the best result for each metric.}
\label{tab:harness}
\centering
\small
\setlength{\tabcolsep}{1.7pt}
\renewcommand{\arraystretch}{1.15}
\begin{tabular*}{\linewidth}{@{\extracolsep{\fill}}lrrrrr@{}}
\toprule
Method & F1$\uparrow$ & TKQR$\uparrow$ & P$\uparrow$ & R$\uparrow$ & FPR$\downarrow$ \\
\midrule
Claude Code
& 9.63 & 6.59 & \textbf{20.86} & 6.26 & \textbf{5.00} \\
OpenHands
& 10.61 & 8.99 & 14.34 & 8.42 & 32.50 \\
\textbf{\model{}}
& \textbf{19.86} & \textbf{18.79} & 19.05 & \textbf{20.73} & \textbf{5.00} \\
\bottomrule
\end{tabular*}
\par\smallskip
\raggedright\footnotesize
FPR is measured on fully specified tasks.
\end{minipage}
\par

The recall gain over Claude Code comes with a small precision decrease, from 20.86\% to 19.05\%.
Its 5.00\% FPR matches Claude Code and is below OpenHands at 32.50\%. Therefore, \model{} identifies more key questions on underspecified tasks without increasing the frequency of unnecessary interaction on fully specified tasks relative to Claude Code.

\subsection{Efficient Clarification and Plugin Deployment}
\label{sec:flash-plugin}

\paragraph{On-Demand Question Qualification.}
To reduce the cost of Question Qualification, we introduce
\model{}-Flash, which qualifies candidates on demand.
Adaptive Question Selection first chooses a candidate from the discovered
questions. The candidate then undergoes Relevance, Resolution, and
Behavioral Verification before it can be asked. Rejected candidates are
discarded and selection continues. This ordering avoids qualifying
candidates that the selector never considers.

\begin{figure}[!htbp]
    \centering
    \includegraphics[width=\textwidth]{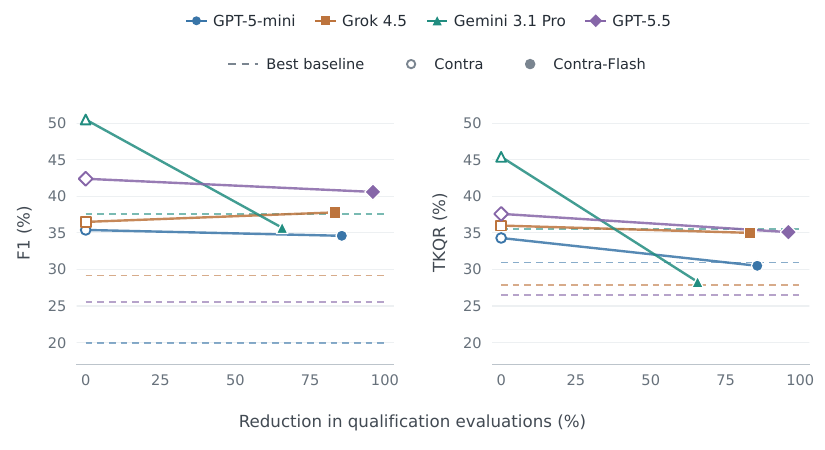}
    \caption{Efficiency and quality of \model{}-Flash. Open/filled markers denote \model{}/Flash. Dashed lines show the best baseline for each agent and metric in Table~\ref{tab:main-results}.}
    \label{fig:flash-tradeoff}
\end{figure}

\paragraph{Efficiency and Clarification Quality.}
Figure~\ref{fig:flash-tradeoff} shows that \model{}-Flash reduces
candidate-level qualification evaluations by 65.7--96.0\%.
F1 remains stable for three of the four coding agents: GPT-5-mini,
Grok~4.5, and GPT-5.5 reduce qualification evaluations by 85.6\%,
83.3\%, and 96.0\%, respectively, while F1 changes by at most
1.8 percentage points. Flash exceeds the best baseline F1 for all three
agents and the best baseline TKQR for Grok~4.5 and GPT-5.5.
For Gemini~3.1 Pro, Flash achieves 35.7\% F1 and 28.3\% TKQR with
65.7\% fewer qualification evaluations, placing its F1 within
1.9 percentage points of the best baseline.
Overall, Flash offers substantial computational savings with stable F1
on most evaluated agents.

\paragraph{Integration with Claude Code.}
We implement \model{} as a Claude Code plugin that reviews decisions after implementation.
The plugin combines the agent's own report with an independent review of
the requirement and code changes to discover unresolved decisions.
It then prioritizes these decisions and explains what the requirement
leaves unspecified, what the agent implemented, and a plausible alternative.
For unresolved implementation choices, Behavioral Verification runs in
the background and uses distinguishing tests to check whether alternative
implementations produce different observable behaviors.
Plugin implementation details appear in Appendix~\ref{app:plugin-implementation}.

\par\addvspace{1.5ex plus 0.5ex minus 0.2ex}
\noindent
\begin{minipage}[t]{0.48\textwidth}
\vspace{0pt}
\textbf{Repository Evaluation.}\quad
We compare Claude Code with and without the plugin on 40 single-function tasks from commits in six Python repositories. Independent review identifies 116 of 262 reported decisions (44.3\%) absent from the agent's own report. We also assess whether the agent's reported decisions explain unspecified information, the implemented choice, and an alternative interpretation.
\end{minipage}\hfill
\begin{minipage}[t]{0.49\textwidth}
\vspace{0pt}
\setlength{\abovecaptionskip}{0pt}
\captionof{table}{Completeness of reported decisions (\%).}
\label{tab:plugin-disclosure}
\centering
\small
\setlength{\tabcolsep}{3.0pt}
\renewcommand{\arraystretch}{1.15}
\begin{tabular*}{\linewidth}{@{\extracolsep{\fill}}lrr@{}}
\toprule
Information reported & Claude Code & +\model{} \\
\midrule
Unspecified information & 4.3 & 59.7 \\
Alternative interpretation & 7.1 & 47.8 \\
\midrule
All three elements & 2.9 & \textbf{46.3} \\
\bottomrule
\end{tabular*}

\par\smallskip
\raggedright\footnotesize
Claude Code and +\model{} report 70 and 67 decisions, respectively.
\end{minipage}
\par

Table~\ref{tab:plugin-disclosure} shows increases from 4.3\% to 59.7\% for unspecified information and from 7.1\% to 47.8\% for alternative interpretations. Reports covering all three elements also rise from 2.9\% to 46.3\%. Therefore, the plugin surfaces unreported decisions and makes unresolved choices explicit. Independent review expands the decisions available for inspection, while the structured report explains why each choice may require confirmation. Together, they help developers assess whether an implemented choice matches their intent. Evaluation details appear in Appendix~\ref{app:plugin-evaluation}.

\section{Conclusion}
\label{sec:conclusion}

We presented \model{}, a training-free method for selective clarification
in coding agents. It combines Question Discovery, Question Qualification,
and Adaptive Question Selection. Semantic checks and execution evidence
identify consequential choices, while interaction history guides which
questions still require user input. Experiments show consistent improvements
in F1 and TKQR across four coding agents and gains over existing coding
harnesses. \model{}-Flash reduces qualification evaluations, while the
Claude Code plugin makes unresolved implementation choices and their
alternatives explicit for developer review. Future work will evaluate how
selective clarification affects developer effort and code quality over
longer development sessions.

\clearpage

\subsection*{AI use statement}

We used generative AI tools to draft, translate, and edit manuscript text, check references, assist with figure design, and generate the illustrative cartoon in Appendix~\ref{app:plugin-walkthrough}. LLM use for program and test generation, clarification evaluation, repository task labeling, and report assessment is described in Sections~\ref{sec:method}--\ref{sec:experiments} and Appendix~\ref{app:plugin}; all 40 repository task labels received manual review. The authors take full responsibility for the final content.

\subsection*{Reproducibility statement}

Section~\ref{sec:method} describes the three stages of \model{}, including program sampling and the behavioral verification criteria. Section~\ref{sec:experimental-setup} specifies the benchmark, baselines, models, question budget, and evaluation protocol, while Section~\ref{sec:downstream-code-generation} describes downstream code evaluation. Appendix~\ref{app:plugin} details the plugin implementation and repository evaluation, including software and model versions, task sampling and its random seed, manual label review, and report-quality assessment.

The released qualification and adaptive-selection code and developer plugin are available at \url{https://github.com/fangz-cs/Contra}. Benchmark data and experiment records are not included in this repository.


\bibliography{references}

\begin{thebibliography}{23}
\providecommand{\natexlab}[1]{#1}
\providecommand{\url}[1]{\texttt{#1}}
\expandafter\ifx\csname urlstyle\endcsname\relax
  \providecommand{\doi}[1]{doi: #1}\else
  \providecommand{\doi}{doi: \begingroup \urlstyle{rm}\Url}\fi

\bibitem[{Anthropic}(2025)]{claude-code}
{Anthropic}.
\newblock {Claude Code}, 2025.
\newblock URL \url{https://github.com/anthropics/claude-code}.
\newblock Software repository. Accessed September 26, 2026.

\bibitem[Edwards \& Schuster(2026)Edwards and Schuster]{ask-or-assume}
Nicholas Edwards and Sebastian Schuster.
\newblock Ask or assume? uncertainty-aware clarification-seeking in coding
  agents.
\newblock \emph{arXiv preprint arXiv:2603.26233}, 2026.
\newblock URL \url{https://arxiv.org/abs/2603.26233}.

\bibitem[Fakhoury et~al.(2024)Fakhoury, Naik, Sakkas, Chakraborty, and
  Lahiri]{fakhoury2024llm}
Sarah Fakhoury, Aaditya Naik, Georgios Sakkas, Saikat Chakraborty, and
  Shuvendu~K. Lahiri.
\newblock {LLM}-based test-driven interactive code generation: User study and
  empirical evaluation.
\newblock \emph{IEEE Transactions on Software Engineering}, 50\penalty0
  (9):\penalty0 2254--2268, 2024.
\newblock \doi{10.1109/TSE.2024.3428972}.

\bibitem[Fang et~al.(2026)Fang, Jin, Dong, Li, Zhang, Jin, and
  Li]{clarifycodebench}
Zheng Fang, Dongming Jin, Yihong Dong, Yongmin Li, Kechi Zhang, Zhi Jin, and
  Ge~Li.
\newblock {ClarifyCodeBench}: Evaluating {LLMs} on clarifying ambiguous
  requirements for code generation.
\newblock \emph{arXiv preprint arXiv:2607.00711}, 2026.
\newblock URL \url{https://arxiv.org/abs/2607.00711}.

\bibitem[{Google DeepMind}(2026)]{gemini31pro}
{Google DeepMind}.
\newblock {Gemini 3.1 Pro} model card, 2026.
\newblock URL \url{https://deepmind.google/models/model-cards/gemini-3-1-pro/}.

\bibitem[Hong et~al.(2024)Hong, Zhuge, Chen, Zheng, Cheng, Wang, Zhang, Wang,
  Yau, Lin, Zhou, Ran, Xiao, Wu, and Schmidhuber]{metagpt}
Sirui Hong, Mingchen Zhuge, Jonathan Chen, Xiawu Zheng, Yuheng Cheng, Jinlin
  Wang, Ceyao Zhang, Zili Wang, Steven Ka~Shing Yau, Zijuan Lin, Liyang Zhou,
  Chenyu Ran, Lingfeng Xiao, Chenglin Wu, and J{\"u}rgen Schmidhuber.
\newblock {MetaGPT}: Meta programming for a multi-agent collaborative
  framework.
\newblock In \emph{International Conference on Learning Representations},
  volume 2024, pp.\  23247--23275, 2024.
\newblock URL \url{https://openreview.net/forum?id=VtmBAGCN7o}.

\bibitem[J{\"a}rvelin \& Kek{\"a}l{\"a}inen(2002)J{\"a}rvelin and
  Kek{\"a}l{\"a}inen]{jarvelin2002cumulated}
Kalervo J{\"a}rvelin and Jaana Kek{\"a}l{\"a}inen.
\newblock Cumulated gain-based evaluation of {IR} techniques.
\newblock \emph{ACM Transactions on Information Systems}, 20\penalty0
  (4):\penalty0 422--446, 2002.
\newblock \doi{10.1145/582415.582418}.

\bibitem[Jia et~al.(2025)Jia, Morris, Ye, Sarro, and Mechtaev]{specfix}
Haoxiang Jia, Robbie Morris, He~Ye, Federica Sarro, and Sergey Mechtaev.
\newblock Automated repair of ambiguous problem descriptions for {LLM}-based
  code generation.
\newblock \emph{arXiv preprint arXiv:2505.07270}, 2025.
\newblock URL \url{https://arxiv.org/abs/2505.07270}.

\bibitem[Kuhn et~al.(2022)Kuhn, Gal, and Farquhar]{clam}
Lorenz Kuhn, Yarin Gal, and Sebastian Farquhar.
\newblock {CLAM}: Selective clarification for ambiguous questions with
  generative language models.
\newblock \emph{arXiv preprint arXiv:2212.07769}, 2022.
\newblock URL \url{https://arxiv.org/abs/2212.07769}.

\bibitem[Li et~al.(2023)Li, Mesgar, Martins, and Gurevych]{codeclarqa}
Haau-Sing~(Xiaocheng) Li, Mohsen Mesgar, Andr{\'e} Martins, and Iryna Gurevych.
\newblock Python code generation by asking clarification questions.
\newblock In \emph{Proceedings of the 61st Annual Meeting of the Association
  for Computational Linguistics (Volume 1: Long Papers)}, pp.\  14287--14306,
  2023.
\newblock \doi{10.18653/v1/2023.acl-long.799}.
\newblock URL \url{https://aclanthology.org/2023.acl-long.799/}.

\bibitem[Mu et~al.(2024)Mu, Shi, Wang, Yu, Zhang, Wang, Liu, and
  Wang]{clarifygpt}
Fangwen Mu, Lin Shi, Song Wang, Zhuohao Yu, Binquan Zhang, Chenxue Wang,
  Shichao Liu, and Qing Wang.
\newblock {ClarifyGPT}: A framework for enhancing {LLM}-based code generation
  via requirements clarification.
\newblock \emph{Proceedings of the ACM on Software Engineering}, 1\penalty0
  (FSE):\penalty0 2332--2354, 2024.
\newblock \doi{10.1145/3660810}.

\bibitem[{OpenAI}(2025)]{gpt5-system-card}
{OpenAI}.
\newblock {GPT-5} system card, 2025.
\newblock URL \url{https://openai.com/index/gpt-5-system-card/}.

\bibitem[{OpenAI}(2026)]{gpt55-system-card}
{OpenAI}.
\newblock {GPT-5.5} system card, 2026.
\newblock URL \url{https://openai.com/index/gpt-5-5-system-card/}.

\bibitem[{Qwen Team}(2026)]{qwen38-27b}
{Qwen Team}.
\newblock {Qwen3.8-27B}, 2026.
\newblock URL \url{https://huggingface.co/Qwen/Qwen3.8-27B}.
\newblock Model card. Accessed September 26, 2026.

\bibitem[{SpaceXAI}(2026)]{grok45}
{SpaceXAI}.
\newblock Introducing {Grok 4.5}, 2026.
\newblock URL \url{https://x.ai/news/grok-4-5}.

\bibitem[Tian \& Chen(2026)Tian and Chen]{specine}
Zhao Tian and Junjie Chen.
\newblock Aligning requirement for large language model's code generation.
\newblock In \emph{Proceedings of the 2026 IEEE/ACM 48th International
  Conference on Software Engineering}, 2026.
\newblock \doi{10.1145/3744916.3764572}.
\newblock URL \url{https://doi.org/10.1145/3744916.3764572}.

\bibitem[Vijayvargiya et~al.(2026{\natexlab{a}})Vijayvargiya, Viswanathan, and
  Neubig]{vijayvargiya2026asking}
Sanidhya Vijayvargiya, Vijay Viswanathan, and Graham Neubig.
\newblock Asking what matters: Reward-driven clarification for software
  engineering tasks.
\newblock \emph{arXiv preprint arXiv:2604.14624}, 2026{\natexlab{a}}.
\newblock URL \url{https://arxiv.org/abs/2604.14624}.

\bibitem[Vijayvargiya et~al.(2026{\natexlab{b}})Vijayvargiya, Zhou, Yerukola,
  Sap, and Neubig]{ambig-swe}
Sanidhya Vijayvargiya, Xuhui Zhou, Akhila Yerukola, Maarten Sap, and Graham
  Neubig.
\newblock {Ambig-SWE}: Interactive agents to overcome underspecificity in
  software engineering.
\newblock In \emph{International Conference on Learning Representations},
  2026{\natexlab{b}}.
\newblock URL \url{https://openreview.net/forum?id=X2yzXtH4wp}.

\bibitem[Wang et~al.(2025)Wang, Li, Song, Xu, Tang, Zhuge, Pan, Song, Li,
  Singh, Tran, Li, Ma, Zheng, Qian, Shao, Muennighoff, Zhang, Hui, Lin,
  Brennan, Peng, Ji, and Neubig]{openhands}
Xingyao Wang, Boxuan Li, Yufan Song, Frank~F. Xu, Xiangru Tang, Mingchen Zhuge,
  Jiayi Pan, Yueqi Song, Bowen Li, Jaskirat Singh, Hoang~H. Tran, Fuqiang Li,
  Ren Ma, Mingzhang Zheng, Bill Qian, Yanjun Shao, Niklas Muennighoff, Yizhe
  Zhang, Binyuan Hui, Junyang Lin, Robert Brennan, Hao Peng, Heng Ji, and
  Graham Neubig.
\newblock {OpenHands}: An open platform for {AI} software developers as
  generalist agents.
\newblock In \emph{International Conference on Learning Representations},
  volume 2025, pp.\  65882--65919, 2025.
\newblock URL \url{https://openreview.net/forum?id=OJd3ayDDoF}.

\bibitem[Wu et~al.(2025)Wu, Chaudhary, Abrahamyan, Khaku, Wei, and
  Fard]{clarifycoder}
Jie~JW Wu, Manav Chaudhary, Davit Abrahamyan, Arhaan Khaku, Anjiang Wei, and
  Fatemeh~H. Fard.
\newblock Can code language models learn clarification-seeking behaviors?
\newblock \emph{arXiv preprint arXiv:2504.16331}, 2025.
\newblock URL \url{https://arxiv.org/abs/2504.16331}.

\bibitem[Yang et~al.(2024)Yang, Jimenez, Wettig, Lieret, Yao, Narasimhan, and
  Press]{swe-agent}
John Yang, Carlos~E. Jimenez, Alexander Wettig, Kilian Lieret, Shunyu Yao,
  Karthik Narasimhan, and Ofir Press.
\newblock {SWE-agent}: Agent-computer interfaces enable automated software
  engineering.
\newblock In \emph{Advances in Neural Information Processing Systems},
  volume~37, pp.\  50528--50652, 2024.
\newblock \doi{10.52202/079017-1601}.

\bibitem[Zhang et~al.(2025)Zhang, Knox, and Choi]{zhang2025modeling}
Michael Zhang, W.~Bradley Knox, and Eunsol Choi.
\newblock Modeling future conversation turns to teach {LLMs} to ask clarifying
  questions.
\newblock In \emph{International Conference on Learning Representations},
  volume 2025, pp.\  60722--60742, 2025.
\newblock URL
  \url{https://proceedings.iclr.cc/paper_files/paper/2025/hash/97e2df4bb8b2f1913657344a693166a2-Abstract-Conference.html}.

\bibitem[Zhang \& Choi(2025)Zhang and Choi]{zhang2025clarify}
Michael~JQ Zhang and Eunsol Choi.
\newblock Clarify when necessary: Resolving ambiguity through interaction with
  {LMs}.
\newblock In \emph{Findings of the Association for Computational Linguistics:
  NAACL 2025}, pp.\  5541--5558, 2025.
\newblock \doi{10.18653/v1/2025.findings-naacl.306}.
\newblock URL \url{https://aclanthology.org/2025.findings-naacl.306/}.

\end{thebibliography}
\bibliographystyle{iclr2027_conference}

\clearpage
\appendix
\section{One Request, Five Decisions: An Illustrative Walkthrough}
\label{app:plugin-walkthrough}

\begin{figure}[!htbp]
    \centering
    {\small\sffamily
    \makebox[0.32\textwidth]{(a) ``Looks good!''}\hfill
    \makebox[0.32\textwidth]{(b) Weeks later: ``Why?''}\hfill
    \makebox[0.32\textwidth]{(c) Review with \model{}}\par}
    \smallskip
    \includegraphics[width=\textwidth]{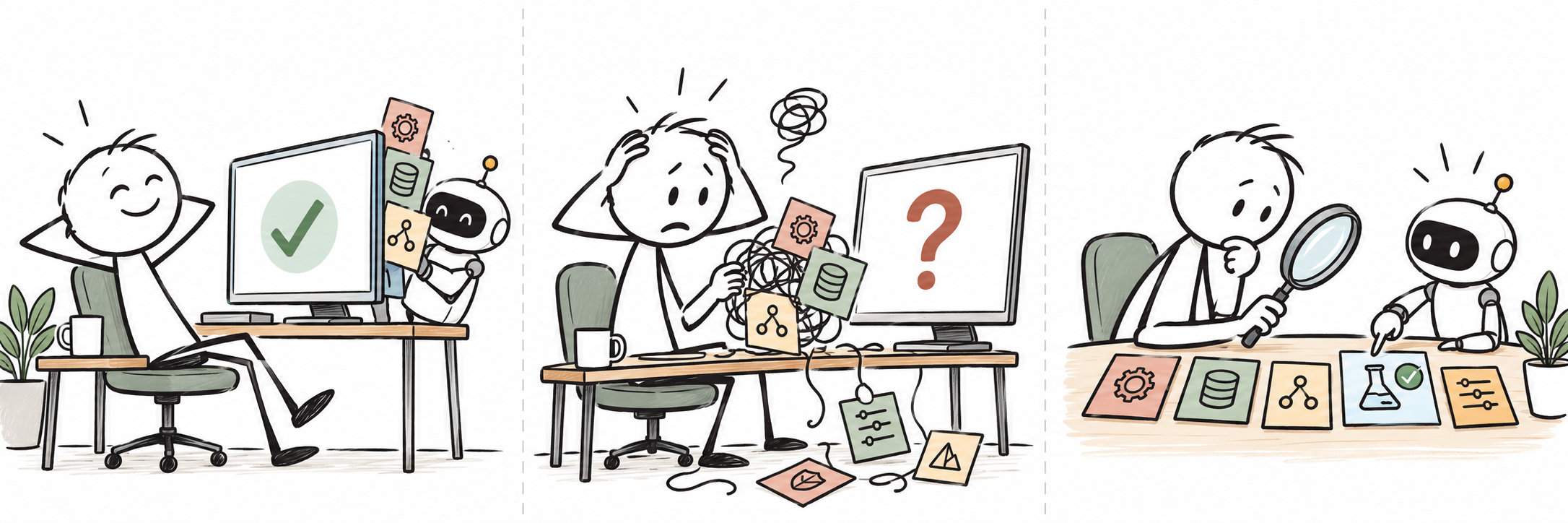}
    \caption{A completion report can leave implementation choices implicit. Those choices may become consequential later. The plugin makes them available for review while the requirement and change are still in context.}
    \label{fig:plugin-story}
\end{figure}

Our plugin demonstration follows a short request in Requests: \emph{``Add Brotli decoding support.''} The recorded change and its five reported decisions show what \model{} exposes. Figure~\ref{fig:plugin-story} extends this example with a hypothetical later user complaint to illustrate why these decisions matter.

\paragraph{A reassuring completion report.}
The agent replaces the fixed \texttt{Accept-Encoding} value \texttt{gzip, deflate} with \texttt{DEFAULT\_ACCEPT\_ENCODING}. Its final reply says that Brotli is included when available and that the Quickstart documentation has been updated. To the developer, the request appears complete. Yet the report does not explain how the new behavior interacts with dependency versions, which encodings the header may advertise, or whether the new constant is intended to become public API.

\paragraph{The choice that surfaces later.}
In the demonstration, Brotli support requires \texttt{urllib3}~1.25 or later, while the package still declares a minimum of 1.21.1. Imagine a user returning three weeks later with version 1.24: the header falls back to the older encodings, although the documentation describes Brotli support. The request never specified whether to raise the dependency bound, warn about unavailable support, or accept the fallback. Each policy has different compatibility implications. The delayed complaint illustrates why a seemingly local edit can leave a consequential decision unresolved.

\paragraph{Five decisions become reviewable.}
With the plugin, the same change is followed by the structured report in Figure~\ref{fig:plugin-interface}. Two entries concern changes beyond the explicit request: introducing a public name in \texttt{requests.utils} and updating the Quickstart documentation. Two concern deferred work: the dependency-bound mismatch and missing regression coverage for the new header. The fifth identifies an implementation choice: forwarding all encodings detected by \texttt{urllib3}, including \texttt{zstd}, instead of adding only \texttt{br}. Together, these entries make scope, deferred work, and alternative interpretations available for review.

\paragraph{From an observation to executable evidence.}
For the fifth entry, the demonstration links a distinguishing test, \path{tests/test_contra_default_headers.py}, showing a behavioral difference between the implemented encoding policy and an alternative. Appendix~\ref{app:plugin-implementation} describes the verification procedure. The developer can inspect this evidence, confirm the intended behavior, and retain the test as a record. The resulting discussion is concrete: should this feature enable Brotli alone, or delegate the advertised encoding set to the dependency?

\clearpage
\section{Plugin Implementation and Evaluation Details}
\label{app:plugin}

\subsection{Plugin Implementation}
\label{app:plugin-implementation}

The Claude Code plugin reviews implementation decisions at the end of an interaction. A hook records the working tree when the user submits a request. When the agent finishes, another hook compares the updated tree with this snapshot to identify changes made during the interaction. This avoids including unrelated changes that were already present before the request.

The plugin then runs two processes in parallel. One extracts decisions from the agent's final report using \texttt{claude-haiku-4-5-20251001}. The other independently reviews the requirement, code changes, and repository context using \texttt{claude-opus-5}, without reading that report. Their outputs are merged using text overlap for entries concerning the same file and decision category. The combined report distinguishes implementation choices, changes beyond the requested scope, and work the agent decided not to perform. Rule-based prioritization uses these categories together with affected scope, trigger conditions, reversibility, and verification outcomes.

For an unresolved implementation choice, the plugin uses \texttt{claude-opus-5} to generate a distinguishing test and a patch implementing an alternative interpretation. It executes the test on the current and patched implementations in isolated repository copies. A test that passes on the current implementation and fails on an executable alternative provides evidence of a behavioral difference. Patch or test-execution errors are distinguished from such evidence. In normal use, verification runs in the background after the initial report. Distinguishing tests can be retained in the repository to document the implemented behavior, and a persistent record tracks decisions across interactions.

\subsection{Repository Evaluation}
\label{app:plugin-evaluation}

The evaluation uses 40 single-function commits from six Python repositories, namely Rich, Flask, Requests, Click, HTTPX, and Typer. The tasks are selected by stratified random sampling, with 20 labeled underspecified and 20 labeled fully specified, using seed 20260903. For each task, the repository is restored to its pre-commit state, and the commit message serves as the requirement. Claude Code is asked to modify the target function and briefly explain the change. The final plugin runs use Claude Code 2.1.261 with \texttt{claude-opus-5} and the research version of the plugin evaluated on September 5, 2026. Each task has one baseline run and one plugin run. Verification runs synchronously during evaluation so that its outcomes are collected before the process exits.

Task labels are generated by \texttt{claude-opus-5} from the commit message and pre-change code, then manually reviewed for all 40 tasks. The criterion is whether this information determines the required observable behavior, without filling missing information from implementation conventions.

Report quality is evaluated from the agent's final reply. We use \texttt{claude-opus-5} to extract decision entries and check whether each states the unspecified information, the implemented choice, and a plausible alternative interpretation. Table~\ref{tab:plugin-disclosure} reports the proportion of entries containing each element and all three together. The denominators are 70 entries without the plugin and 67 with it.

The decision-discovery analysis uses the plugin's combined structured output, a separate collection from the final-reply entries above. Among its 262 decisions, 116 appear only in independent review, 89 only in the agent-derived list, and 57 in both. The repository evaluation thus measures decision discovery and report completeness.

\clearpage
\section{Plugin Interface and Operational Scope}
\label{app:plugin-interface}

\begin{figure}[!htbp]
    \centering
    \includegraphics[width=\textwidth]{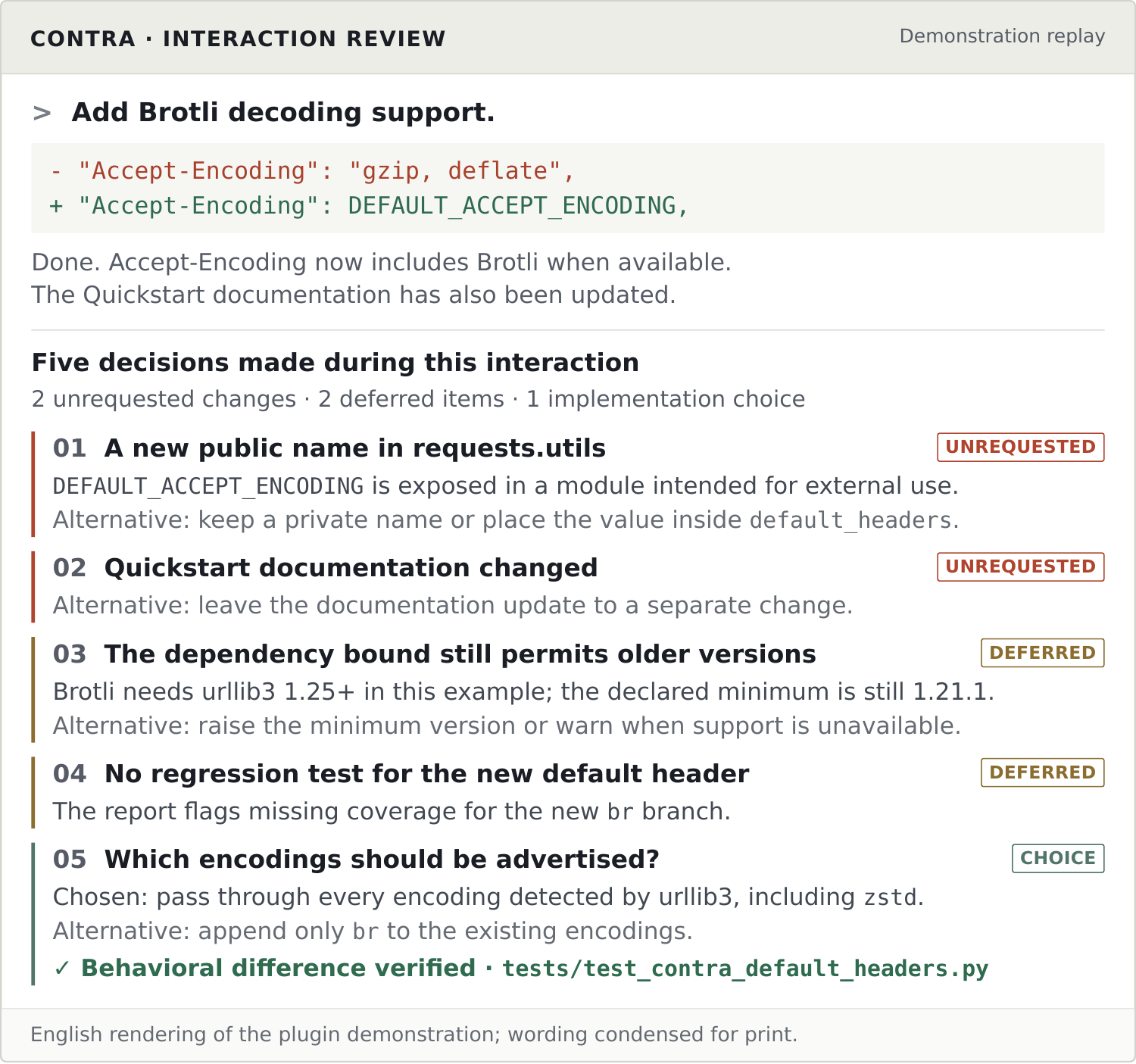}
    \caption{English rendering of the supplied plugin demonstration, condensed for print. It preserves the five decisions, three categories, and test-artifact path. The verification indicator applies to the fifth entry.}
    \label{fig:plugin-interface}
\end{figure}

Figure~\ref{fig:plugin-interface} presents an English rendering of the demonstration. The request, code change, and completion report remain visible above the five decisions, connecting each concern to the interaction that produced it.

\paragraph{Reading the report.}
An entry exposes the implemented choice or deferred work and, where available, an alternative. The developer can accept the choice, request a revision, or defer it explicitly. A separate verification indicator identifies entries supported by an observed behavioral difference.

\paragraph{Cost and applicability.}
This single-run demonstration reports 40\,s (\$0.47) for coding, 90\,s (\$0.17) for reporting, and 15\,s (\$0.06) for background verification. The report appears about 130\,s after the request; the total reported cost, including verification, is \$0.70. In this build, executable evidence uses Python and \texttt{pytest}; other supported source languages receive decision reports. Verification supports the behavior exercised by each distinguishing test.

\end{document}